\documentclass[11pt]{article}
\usepackage{amsmath}
\usepackage{amsfonts}
\usepackage{url}
\usepackage{color}
\usepackage[square]{natbib}
\usepackage{algpseudocode}
\usepackage{algorithm}
\usepackage{listings}
\usepackage{caption}
\usepackage{booktabs}
\usepackage{afterpage}
\definecolor{NavyBlue}{cmyk}{0.94,0.54,0,0}
\usepackage[raiselinks=true,%
        colorlinks=true,%
        linkcolor=NavyBlue,%
        menucolor=NavyBlue,%
        citecolor=NavyBlue,%
        urlcolor=NavyBlue,%
        bookmarks=true,%
        bookmarksopenlevel=1,%
        bookmarksopen=true,%
        bookmarksnumbered=true,%
        hyperindex=true,%
        plainpages=false,%
        pdfpagelabels=true,%
        ]{hyperref}

\DeclareCaptionFormat{mylst}{\hrule#1#2#3}
\newcommand{\alphab}{\mathcal{A}}

\begin{document}
\setcitestyle{numbers}

\title{Interleaved entropy coders}
\author{Fabian Giesen}
\maketitle

\begin{abstract}
The ANS family of arithmetic coders developed by Jarek Duda has the
unique property that encoder and decoder are completely symmetric in the
sense that a decoder reading bits will be in the exact same state that
the encoder was in when writing those bits---all ``buffering'' of
information is explicitly part of the coder state and identical between
encoder and decoder. As a consequence, the output from multiple ABS/ANS
coders can be interleaved into the same bitstream without any additional
metadata. This allows for very efficient encoding and decoding on CPUs
supporting superscalar execution or SIMD instructions, as well as GPU
implementations. We also show how interleaving without additional
metadata can be implemented for \emph{any} entropy coder, at some
increase in encoder complexity.
\end{abstract}

\section{Introduction}

Duda's recent paper~\cite{DudaANS} describes a family of entropy coders
based on what he calls ``Asymmetric numeral systems'' (ANS for short).
They are closely related to conventional arithmetic coders~\cite{moffat1998arithmetic}
or ``range coders''~\cite{martin1979range} and have very similar coding
performance. However, the underlying construction is quite different:
classical arithmetic coders conceptually view the bitstream as encoding
a dyadic fraction in $[0,1)$, and proceed by a process of interval
subdivisions. ``Infinite-precision'' ANS encoders, by contrast, view the
bitstream as a single integer $x \in \mathbb{N}$ that encodes the entire
message, where the coded symbols are derived from the value of the remainder
$x \bmod m$ for some integer $m$. Finally, ``streaming'' ANS encoders
perform the same process using fixed-width integer arithmetic ($x < 2^p$
for some integer $p$; often $p=32$ or $p=64$ for machines with 32- or
64-bit word sizes, respectively), which allows for efficient implementations.

Streaming ANS coders are based on a fairly unique construction that
ensures that a decoder reading bits will always be in the exact same state
the corresponding encoder was in while writing these bits; all buffering
of ``not yet final'' bits is completely identical between encoder and decoder.
This property makes it possible to run multiple encoders concurrently and
interleave them into a single bit stream in a way that can be reconstructed
on the decoder side without additional metadata. The decoder can likewise
read these multiple interleaved streams concurrently, provided the streams
are independent. The resulting algorithms are both simple and efficient,
and can be vectorized given a suitable instruction set.

We will first review ANS coders and their unique construction. It will be
shown how they can be interleaved easily, and how this property can be
exploited for SIMD and GPU applications.

\section{Streaming ANS coding}

Let $\alphab := \{ 0, 1, \dots, n-1 \}$ be a finite symbol alphabet.
Furthermore, let $\{p_s\}_{s \in \alphab}$ be the probabilities of the
respective symbols, $\sum_{s \in \alphab} p_s=1$.
In ANS, the encoding and decoding processes revolve around a pair
of functions
\[ C : \alphab \times \mathbb{N} \rightarrow \mathbb{N} \]
and
\[ D : \mathbb{N} \rightarrow \alphab \times \mathbb{N}. \]
$C$, the coding function, takes a symbol $s$ and the current state $x$
and produces a new value $x' = C(s,x)$ that encodes both the new symbol $s$
and the original state $x$. $D$, the decoding function,
is $C$'s inverse: $D(x') = (s,x)$ and $C(D(x)) = x$. Different choices of
$C$ and $D$ correspond to different coder variants; \cite{DudaANS}~covers
a wide variety of options with different constraints on the alphabet and
probability distributions, and different trade-offs of memory usage vs.\@
computational complexity.

By construction, $x$ acts like a stack: the last value encoded will be first
value returned by the encoder. This is a direct result of $D$ being the
inverse of $C$, since for example:
\[ D(C(b, C(a, x))) = (b, C(a,x)) \]
The same also holds for the encoded bit stream: the last bits
written by the encoder will be consumed first by the decoder. This article
will use the convention that the encoder processes symbols backwards, from
last to first, and also writes the bit stream backwards, while the decoder
runs forwards.

As more and more symbols get encoded into $x$, its value will, eventually,
grow beyond all bounds. But for an efficient implementation, we
must have fixed bounds on the size of $x$; usually $x < 2^p$ for some $p$.
To this end, we define a ``normalized interval''
\[ I := \{ L, L+1, \dots, bL - 1 \} \]
where $L$ is an arbitrary positive integer and $b$ an integer $\ge 2$
that defines the ``radix'' of the coder: a coder with $b=2$ emits
individual bits, one with $b=256$ writes 8-bit bytes at a time, and so forth.
We then require that both encoder and decoder maintain the invariant that
$x \in I$, by writing some of the bits from $x$ and reducing it when it gets
too large in the encoder, or reading additional bits when it gets too small
in the decoder. Algorithm~\ref{alg:sans} illustrates the general idea. Note
that the encoder is the exact inverse of the decoder: it performs the inverses
of the individual decoder steps, in reverse order. The final state of the encoder
(which corresponds to the state after coding the \emph{first} symbol in the message,
since the encoder iterates over the message backwards) needs to be transmitted along with
the rest of the bitstream so the decoder knows what state to start in.

\begin{algorithm}[tb]
\caption{Streaming ANS decoder/encoder}
\label{alg:sans}
\begin{minipage}{0.48\textwidth}
\begin{algorithmic}
        \Require{$x \in I$}
        \Ensure{$x \in I$}
        \Function{Decode}{iobuf}
                \State $s, x \gets D(x)$
                \While{$x \not\in I$}
                        \State $x \gets bx + \textrm{iobuf.read()}$
                \EndWhile
                \State \Return $s$
        \EndFunction
\end{algorithmic}
\end{minipage}
\begin{minipage}{0.48\textwidth}
\begin{algorithmic}
        \Require{$x \in I$}
        \Ensure{$x \in I$}
        \Procedure{Encode}{$s$,iobuf}
                \While{$C(s,x) \not\in I$}
                        \State iobuf.emit$(x \bmod b)$
                        \State $x \gets \lfloor x / b \rfloor$
                \EndWhile
                \State $x \gets C(s,x)$
        \EndProcedure
\end{algorithmic}
\end{minipage}
\end{algorithm}

It is not at all clear that this approach works, and in fact given an arbitrarily
chosen pair of mutually inverse $C$ and $D$ it usually will not. For example,
given an arbitrary $C$, there is no guarantee that the loop in \textsc{Encode}
will terminate, or that encoder and decoder will always go through the same sequence
of states $x$. However, it can be shown that if the \emph{precursor sets}
\[ I_s := \{ x \: | \: C(s,x) \in I \} \]
are of the form $I_s = \{ k, k+1, \dots, bk-1 \}, k \ge 1$---a property that
Duda calls ``$b$-uniqueness'' in~\cite{DudaANS}---the encoder and decoder in
algorithm~\ref{alg:sans} will stay synchronized. Duda gives several general
constructions (uABS, rANS/rABS, tANS/tABS) that are guaranteed
to produce valid $C$ for different types of alphabets and probability distributions.
For the purposes of this paper, it does not matter which particular $C$ is chosen.

\subsection{Example}

For a concrete example, suppose we wish to code a message from
the two-symbol alphabet $\alphab = \{ a, b \}$ with $p_a = 1/4$,
$p_b = 3/4$. We define:
\begin{align*}
C(a,x) &= 4x \\
C(b,x) &= 4 \lfloor x/3 \rfloor + (x \bmod 3) + 1
\end{align*}
Note that $a$ gets sent to values that are divisible by 4, whereas
$b$ gets sent to values that are not. Furthermore, in both cases,
the original value of $x$ is easy to recover from the new value,
and we have the straightforward inverse:
\[
D(x) = \begin{cases}
        (a, x/4) & \text{if $x \equiv 0 \mod 4$,} \\
        (b, 3 \lfloor x/4 \rfloor + (x \bmod 4) - 1) & \text{otherwise.}
\end{cases}
\]
When an $a$ is coded using $C$, $x$ grows by a factor of $4=1/p_a$; when a
$b$ is coded, $x$ grows by a factor of approximately $4/3=1/p_b$. More
generally, in an ANS coder, coding the symbol $s$ will grow $x$ by
a factor of approximately $1/p_s$. This is a rough equivalent to
the size of the range in a regular arithmetic coder, which \emph{shrinks}
by a factor of approximately $p_s$ for every coded symbol.

Now suppose we define $L=16$ and $b=2$, making our normalization interval
$I = \{ 16, 17, \dots, 31 \}$. Some computation shows that $a$ and $b$'s
precursor sets are given by
\begin{align*}
I_a &= \{ 4, 5, 6, 7 \} \\
I_b &= \{ 12, 13, \dots, 23 \}
\end{align*}
and both of them are $b$-unique. Therefore, algorithm~\ref{alg:sans}
is guaranteed to work. Figure~\ref{fig:exam} on page~\pageref{fig:exam}
shows a worked-through example of encoding, then decoding, the message
``babba'' with this choice of parameters. Note that encoder and decoder
go through the same sequence of states, just in opposite order. This is
true by construction, since encoder and decoder are designed as exact
inverses of each other. The rest of this paper describes several ways
to exploit this fact.

\begin{figure*}[!t]
\centering
\begin{tabular}{ccc}
\toprule
Encoder & x (state) & Decoder \\
\midrule
\input{exam.txt}
\bottomrule
\end{tabular}
\vskip1em
$L=16$, $b=2$, $I = \{ 16, 17, \dots, 31 \}$
\begin{align*}
C(a,x) &= 4x & I_a = \{ 4, 5, 6, 7 \} \\
C(b,x) &= 4 \lfloor x/3 \rfloor + (x \bmod 3) + 1 & I_b = \{ 12, 13, \dots, 23 \} \\
\end{align*}
\[
D(x) = \begin{cases}
        (a, x/4) & \text{if $x \equiv 0 \mod 4$,} \\
        (b, 3 \lfloor x/4 \rfloor + (x \bmod 4) - 1) & \text{otherwise.}
\end{cases}
\]
\caption{ANS example: Coding ``babba'' with $p(a)=1/4$, $p(b)=3/4$.
Encoder proceeds from bottom to top, decoder from top
to bottom (as indicated). Both go through the exact same sequence of
states and perform I/O in the same places.}
\label{fig:exam}
\end{figure*}
\afterpage{\clearpage}

\section{Interleaving ANS coders}

Streaming ANS, as constructed above (and illustrated in the example), has the
rather unique property that decoder and encoder perform the same sequence of
state transitions and do IO operations at the exact same time (except for the
encoder doing it all in reverse order, that is).

The decoder does not only produce the original message that was passed into
the encoder; it produces it by ``unwinding'' all operations performed by the
encoder, one by one. After decoding the $i$'th symbol $s_i$ in the message,
the decoder ``winds back time'' to the state the encoder was in right before
it encoded symbol $s_{i+1}$.

Thus, encoder and decoder are always in lockstep. Regular arithmetic coders
do not have this property: the decoder is always ahead of the encoder,
in the sense that the decoder's state while decoding $s_i$ will already contain
bits that the encoder only sent after $s_i$ was encoded---sometimes, much later.
The resulting asymmetry makes it hard to mix arithmetic-coded with non-arithmetic-coded
data in the same bit stream without performing an expensive (in terms of
rate) flush operation---motivating designs like the ``bypass coding
mode'' in CABAC~\cite{marpe2003context}, a special fast path in a binary
arithmetic coder to accelerate encoding (and decoding) of near-equiprobable
binary symbols.

With ANS, this is not an issue. Because the decoder and encoder proceed in lockstep,
an encoder can just write raw bits (bytes, \dots) into the output bitstream whenever
it makes sense; the decoder will be at the same position in the bitstream at the same
time and there is no need for an explicit bypass coding mechanism. For example, suppose
that in figure~\ref{fig:exam}, we have four different ``subtypes'' of symbol $a$ called
$a_0$ through $a_3$, all of which are equally likely. In a regular arithmetic coder, we
would need to use a bypass mechanism to encode the two bits denoting which of the four
types it is. In an ANS coder, we can just agree that after reading an $a$, the decoder
will read two additional bits that denote the subtype---and likewise, that \emph{before}
encoding an $a$, the encoder sends the subtype number using two bits.

Suppose we do just this in figure~\ref{fig:exam}, for the first $a$ in the message. After
decoding $a$, the decoder is in state 30---the same state the encoder was in before
encoding the $a$. Writing and reading our two extra bits as just described
corresponds to inserting an extra row right below state 30: the encoder emits two bits,
and the decoder reads two bits. The two sides are still perfectly symmetric; unique
decodability is still guaranteed.

By the same argument, a bitstream can interleave not just an ANS coder with a
``raw'' binary coder, but also multiple ANS coders---or, in fact, any mixture of ANS
and raw binary coders. By ``interleave'', I mean that there are two or more distinct
decoders with distinct states reading from (or likewise, two or more encoders with
distinct states writing to) the same buffer:
\begin{algorithmic}
        \State $s_1 \gets$ coder1.\textsc{Decode}(iobuf)
        \State $s_2 \gets$ coder2.\textsc{Decode}(iobuf)
\end{algorithmic}
No additional metadata is necessary for this, provided that the sequence of
coders and models used is the same between encoder and decoder.

The advantage of using distinct coders (with distinct states) over just having
different models is that they are truly independent and can be processed concurrently,
provided that their probability distributions are either static or evolve
independently. This allows for faster implementations on superscalar processors,
and, once we interleave a larger number of streams---say somewhere between 4--64---also
enables use of SIMD instructions or even performing entropy coding on a GPU.
We will cover both these use cases shortly.

Interleaving multiple streams is not a panacea; the ``independent model
evolution'' requirement precludes certain kinds of context models. It is,
however, very interesting for applications such as image and video coding
that deal with large amounts of data that have a homogeneous structure (e.g.
transform coefficients) and are generally coded using independent contexts
anyway.

\section{Interleaving arbitrary entropy coders}

Note that \emph{any} entropy coder can, without modification,
support the less constrained scenario
\begin{algorithmic}
        \State $s_1 \gets$ coder1.\textsc{Decode}(iobuf1)
        \State $s_2 \gets$ coder2.\textsc{Decode}(iobuf2)
\end{algorithmic}
where ``iobuf1'' and ``iobuf2'' correspond to two physically distinct streams.
However, this is not quite equivalent: say we want to produce a single output,
like a file or a TCP stream. Multiplexing these multiple streams into a
single container requires extra work, and it's not obvious that this can be
done without additional metadata at all.

But in fact, the design of ANS shows precisely how an encoder can be modified
to support such ``free'' interleaving. The key to the normalization procedure
in algorithm~\ref{alg:sans} is that the encoder knows precisely how the
decoder will react to any given input.

We can do the same thing with \emph{any} entropy coder: it may not be
practical to make the encoder directly aware of exactly what the decoder will
do, but luckily this is not necessary---the encoder can just \emph{run} 
(simulate) the decoder to figure out what the actual sequence of reads will be.

To elaborate: suppose we have a multi-stream encoder as above, writing data
into multiple buffers, one per stream. Then, to produce the final bitstream,
the encoder runs an instrumented version of the decoder. For simplicity, suppose
we have two streams, one of which receives the symbols at even positions, with
the other one receiving the symbols at odd positions. Then we run the
instrumented decoder:

\begin{samepage}\begin{algorithmic}
        \State $s_1 \gets$ coder1.\textsc{DecodeInstrumented}(iobuf1, iobuf\_mux)
        \State $s_2 \gets$ coder2.\textsc{DecodeInstrumented}(iobuf2, iobuf\_mux)
\end{algorithmic}\end{samepage}

Here, \textsc{DecodeInstrumented} is just a slightly modified version of
\textsc{Decode} (for whatever entropy coder is used): it reads from the
stream passed in as its first argument, but whenever it reads something,
it immediately writes that value to the buffer given as the second
argument.

Once the instrumented decoder finishes, iobuf\_mux contains the bits from all
streams, in exactly the order that the decoder on the receiving end
will be trying to read them. Moreover, it is not necessary to wait for all
streams to ``finish'' before starting the instrumented decoder; it can run
along with the encoders as a coroutine, and one might hope that doing so
can reduce the memory requirements to a constant amount of storage per stream.
However, there is a small catch; suppose the encoder does something like:

\begin{algorithmic}
        \State coder1.\textsc{Encode}($s_0$, iobuf1)
        \For{$i \gets 1$ to 1000000}
                \State coder2.\textsc{Encode}($s_i$, iobuf2)
        \EndFor
        \State coder1.\textsc{Flush}(iobuf1)
        \State coder2.\textsc{Flush}(iobuf2)
\end{algorithmic}

In this case, it is highly likely that an instrumented decoder running as a
coroutine will get stuck reading the very first symbol, right until the very
end when coder1 calls \textsc{Flush}; thus, the instrumented decoder will not
actually make any forward progress until the entirety of ``iobuf2'' has been
written, and the encoder will require enough memory to buffer all of it. This
can happen whenever the output rates of different streams are highly
un-balanced.

We can reduce these memory requirements to a predictable amount by requiring
flushes periodically; say we flush at least once every million symbols. This
resolves the problem and guarantees bounded memory usage, at the cost of
some increase in bit rate.

This approach is fully general and works with any entropy coder; that said,
encoding to multiple streams (and then later running the instrumented decoder
to produce the final bit stream) adds extra buffering and complexity to the
encoder; worst-case memory usage is proportional to the size of the output data.
The situation with ANS is similar: Its reverse encoding requirement means
that in practice, the \emph{input} streams will have to be buffered in some way,
at (generally) even higher cost in memory than the output streams would be.
However, interleaving ANS streams is simple on both the encoder and decoder sides;
this simplicity enables efficient vectorized implementations.

\section{SIMD implementation}

Algorithm~\ref{alg:sans} is written as a serial program; however, like any
serial program, it can be systematically converted into vectorized
form: integer arithmetic turns into arithmetic on vectors of integers, memory
loads and stores turn into gather/scatter operations, and control flow can be
turned into SIMD data flow by using predication; see e.g.~\cite{karrenberg2011whole}
for details.

This (entirely mechanical) process is guaranteed to preserve the meaning of
the original program if (and only if) there are no data dependencies between the
computations running simultaneously in different SIMD lanes.

Assuming that static models are used, $C$ and $D$ are pure functions and
free of side-effects; they can be vectorized safely. For adaptive models, they
can still be vectorized safely as long as no two SIMD lanes ever try to update
the same model at the same time; one way to ensure this is to keep separate
contexts for every SIMD lane.

Once $C$ and $D$ are taken care of, the only remaining problem is how to
implement the \texttt{emit} and \texttt{read} operations in the encoder and
decoder, respectively.

For concreteness, suppose $b=2^8=256$ (i.e. byte-aligned IO) and that the
target SIMD width is $N=4$. Listing~\ref{lst:sans_simd} illustrates how such a
decoder can be implemented.

\begin{lstlisting}[caption={SIMD decoder with byte-wise normalization},float=t,label={lst:sans_simd}]
// Decoder state
Vec4_U32 x;
uint8* read_ptr;

Vec4_U32 Decode()
{
  Vec4_U32 s;

  // Decoding function (uABS, rANS etc.).
  s, x = D(x);

  // Renormalization
  // L = normalization interval lower bound.
  // NOTE: not quite equivalent to scalar version!
  // More details in the text.
  for (;;) {
    Vec4_U32 new_bytes;
    Vec4_bool too_small = lessThan(x, L);
    if (!any(too_small))
      break;

    new_bytes = packed_load_U8(read_ptr, too_small);
    x = select(x, (x << 8) | new_bytes, too_small);
    read_ptr += count_true(too_small);
  }

  return s;
}
\end{lstlisting}

This code uses the following utility functions:
\begin{itemize}
\item \texttt{x = select(a, b, cond)} is a SIMD version of C's ternary
  operator: $\texttt{x}_i = \texttt{b}_i$ if $\texttt{cond}_i = \texttt{true}$,
  $\texttt{x}_i = \texttt{a}_i$ otherwise.
\item \texttt{count\_true(x)} returns how many of the lanes of \texttt{x}
  are \texttt{true}.
\item \texttt{packed\_load\_U8(ptr, mask)} performs byte loads from increasing
  addresses for all lanes where \texttt{mask} is \texttt{true}; where \texttt{mask}
  is \texttt{false}, the return value is zero and the load address is not incremented.
  For example, suppose that $\texttt{mask} = \{ \texttt{true},
  \texttt{false}, \texttt{true}, \texttt{true} \}$; then the packed load
  would return $\{ \texttt{ptr[0]}, 0, \texttt{ptr[1]}, \texttt{ptr[2]} \}$.
\end{itemize}

The details of how to map these pseudo-instructions (\texttt{packed\_load} in
particular, ``borrowed'' from Intel's ISPC compiler~\cite{pharr2012ispc}) to
an efficient sequence of real instructions are architecture-dependent. The
example implementation benchmarked in section~\ref{sec:eval} uses SSE 4.1
instructions running on x86; the technique used in this implementation
(building a bit mask denoting which lanes are active and then using a
mask-dependent shuffle on the results of an unaligned memory load) is
fairly universal.

As noted in the listing, this SIMD decoder is not fully equivalent to having
$N$ interleaved scalar decoders reading from the same bitstream: suppose that
at least one of the SIMD lanes needs two or more renormalization steps---that
is, it reads more than one byte from the encoded bitstream.

The regular decoding algorithm will read all bytes for a single stream before
it moves on to decoding the next one. That is, suppose that $N=2$, with lane~0
reading two bytes and lane~1 reading only one byte. The regular encoder from
algorithm~\ref{alg:sans} will first write both bytes for stream~0, followed by
the byte for stream~1---depth-first order, so to speak.

By contrast, the SIMD decoder in listing~\ref{lst:sans_simd} expects
breadth-first order: the first byte from stream~0, followed by the first byte
for stream~1, followed by the second byte from stream~0.

It is possible to account for this on the encoder side, by either using a
corresponding SIMD encoder that emits bytes in the same way, or by performing
the breadth-first traversal ``by hand'' in a scalar encoder. Both strategies work,
but have the unfortunate consequence that the target SIMD width $N$ is now
effectively baked into the bit stream (since it determines how exactly bytes get
interleaved).

An alternative solution is to simply require that no lane ever execute more
than one iteration of the normalization loop per symbol---thus turning the loop
into a simple \texttt{if} statement. This places some requirements on the
model probabilities; for example, with rANS, we can guarantee that no symbol
will ever need to go through more than one normalization iteration when
$b \ge m$, where $b$ is the radix of the encoder and $m$ is the least common
denominator of all symbol probabilities $p_s$. With $b=256$ as in the example
listing, this is impractical for typical alphabet sizes; forcing $m=256$ would mean
that a 256-symbol alphabet could only use a uniform distribution! However, for an
encoder working in 32-bit integers, we can choose $b=2^{16}$ while retaining a
decent probability resolution of 12--14 bits.

The big advantage of choosing our coding parameters such that one iteration of
the normalization loop is always sufficient is that the resulting bitstreams are
exactly the same between the regular serial version and the SIMD variants; it is
not necessary to encode the bitstream in any special way, and the result is not
tied to any particular SIMD width $N$. Thus, this is the option used in the
example code discussed in section~\ref{sec:eval}.

This section covers only the decoder; because ANS decoder and encoder are
symmetrical, we can use the same technique on the encoder as well, this time using
a \texttt{packed\_store} instead of a \texttt{packed\_load} operation.

\section{GPU implementation}

Although the typical \emph{programming model} for GPUs is quite different from
SIMD instructions on general-purpose CPUs, the underlying hardware is not; GPUs
are, in essence, wide SIMD processors with built-in support for predication and
relatively high-performance gather/scatter operations.

As a result, the SIMD approach described above is suitable for use on GPUs too,
and the translation is, for the most part, quite straightforward. However, just
as with the regular SIMD implementation, we need to come up with an efficient
strategy to perform ``packed'' loads and stores.

The underlying idea is to determine in advance which invocations (in GLSL~\cite{glsl430}
parlance) are going to perform a renormalization step; if all invocations know
which invocations are going to perform a read, the packed load (or packed store)
offset computation reduces to a prefix sum per invocation, which can be done
very efficiently. Listing~\ref{lst:sans_gpu} illustrates the idea; this time,
the decoder uses the loop-free approach with $b=2^{16}$ discussed in the previous
section (a looping version would loop over the entire renormalization code and
break once \texttt{renorm\_mask} becomes zero).

\begin{lstlisting}[caption={GPU decoder with 16-bit normalization},float,label={lst:sans_gpu}]
// Decoder state
uint x;
uint16 in_buf[];
uint read_pos;
uint invoc_id; // ID for this invocation

uint Decode()
{
  // Decoding function (uABS, rANS etc.).
  uint s;
  s, x = D(x);

  // Renormalization
  uint invoc_bit = 1 << invoc_id;
  uint renorm_mask = ballot(x < L);
  uint offs = bitCount(renorm_mask & (invoc_bit-1));
  if (x < L)
    x = (x << 16) | in_buf[read_pos + offs];
  read_pos += bitCount(renorm_mask);

  return s;
}
\end{lstlisting}

The listing assumes a work group size of 32 invocations (or smaller), so that
a 32-bit integer is sufficient to hold \texttt{renorm\_mask}. It uses
the CUDA~\cite{nvidia2007compute} \texttt{ballot} instruction to compute
\texttt{renorm\_mask}, which is used both for the prefix sum offset
computation and to advance the read pointer afterwards. \texttt{ballot}
evaluates the given condition on all threads in a warp and returns a bit
mask that has a $1$ bit in the $i$'th position if the condition was true
on thread $i$. When a \texttt{ballot} instruction is not available (or
in case the work groups are larger than the warp size / hardware SIMD
width), its functionality can be synthesized---at some cost in
efficiency---using atomic operations in group shared memory, as illustrated
by listing~\ref{lst:no_ballot}; similarly, larger work groups might require
using 64-bit integers (or arrays of integers) to store \texttt{renorm\_mask},
but the general approach remains the same.

\begin{lstlisting}[caption={Forming \texttt{renorm\_mask} without \texttt{ballot}},float,label={lst:no_ballot}]
shared uint renorm_mask;

// Form renorm_mask
if (x < L)
  atomicOr(renorm_mask, invoc_bit);
else
  atomicAnd(renorm_mask, ~invoc_bit);

// Make sure all invocations finish modifying
// renorm_mask.
groupMemoryBarrier();
barrier();
\end{lstlisting}

No matter which ``ballot'' approach is used, this results in a fully
vectorized entropy coder, with predictable and well-coalescable
memory access patterns, using a small number of GPU registers and
shared memory space---and therefore well-suited for integration into
other decompression kernels, if so desired.

As with the SIMD version, an encoder can be designed using the same
technique, and the bitstreams are fully compatible between interleaved
sca\-lar, CPU SIMD, and GPU implementations, as long as the version with
at most one renormalization step per symbol is used.

\section{Evaluation}
\label{sec:eval}

An implementation of rANS (one of the coders belonging to the ANS
family) that shows 2-way interleaving and SIMD (SSE 4.1) decoding
of an 8-way interleaved rANS stream is available at
\url{https://github.com/rygorous/ryg\_rans}. This version implements
a simple order-0 model with static probability distribution (determined
once per file).

Table~\ref{tab:perf} shows the resulting decompression rates on
an Intel Core i7-2600K CPU (3.4GHz) on various test files from the
Calgary Corpus~\cite{bell1989modeling} (book1, book2, pic), the
Canterbury Corpus~\cite{arnold1997corpus} (E.coli, world192.txt), and
finally the source code for the Linux kernel version 3.14-rc1.

As is evident from the table, 2-way interleaving consistently
achieves speed-ups of $1.6\times$ or more over the non-interleaved
decoder in this test, despite there being no significant difference
in the number of operations executed per symbol. The reason
is that the interleaved decoder benefits greatly from superscalar
and out-of-order execution, while the non-interleaved version has
a long chain of dependent operations and thus can't utilize most
of the CPU's available execution resources.

The 8-way interleaved SIMD version is faster still (despite only using
4-wide SSE4.1 instructions, the code uses 8-way interleaving to, again,
benefit from superscalar execution), although its performance is somewhat
limited by the lack of a fast SIMD ``gather'' instruction that has to be
simulated using scalar loads. Additionally, the SIMD version is
entirely free of branches in the decoder, resulting in performance that
is almost completely independent of the data being processed; compare the
steady performance of the SIMD version against the much more variable
throughput for the scalar versions.

\begin{table}
\centering
\begin{tabular}{lrrrrr}
\toprule
                   & serial & \multicolumn{2}{c}{2-way} & \multicolumn{2}{c}{8-way, SIMD} \\
File               & MiB/s & MiB/s & speed-up & MiB/s & speed-up \\
\midrule
book1              & 219.2 & 364.8 & 1.66 & 565.6 & 2.58 \\
book2              & 202.4 & 355.4 & 1.76 & 565.6 & 2.79 \\
pic                & 244.9 & 445.6 & 1.82 & 565.4 & 2.31 \\
E.coli             & 265.9 & 500.6 & 1.88 & 570.8 & 2.15 \\
world192.txt       & 209.0 & 337.7 & 1.62 & 564.3 & 2.70 \\
linux-3.14-rc1.tar & 217.6 & 359.0 & 1.65 & 573.2 & 2.63 \\
\bottomrule
\end{tabular}
\caption{Decompression performance of differently interleaved rANS decoders
on various test files (all speed-ups relative to non-interleaved serial decoder).}
\label{tab:perf}
\end{table}

\section{Conclusion}

The ANS family of coders has the unique advantage over regular arithmetic
coders that all buffering of information is identical between encoder
and decoder; the timing of input/output operations only depends on the
coder's state variable $x$, and both encoder and decoder go through the
same sequence of states and perform IO at the same time.

Consequently, ANS coders can easily support efficient bypass coding as
well as interleaving of data from multiple encoders without any additional
metadata. Inspired by ANS, we show how the same property can be achieved
with any entropy coder, by (conceptually) running an instrumented version
of a ``multi-stream'' decoder at encode time.

Interleaving multiple independent ANS coders enable both much
faster scalar entropy coders (with speed-ups of over $1.6\times$ compared
to the baseline), and SIMD implementations (with speed-up factors above
$2\times$). The same technique is easily adapted to GPUs.

\section*{Acknowledgments}

Thanks to my colleagues Charles~Bloom and Sean~Barrett for reviewing
earlier drafts of this paper and making valuable suggestions.

\bibliographystyle{plainnat}
\bibliography{interleaved_entropy}

\end{document}